\documentclass[twocolumn,showpacs,preprintnumbers,amsmath,amssymb,nofootinbib]{revtex4}
\usepackage{graphicx}
\usepackage{epsfig}
\usepackage{dcolumn}
\usepackage{bm}

\def\be{\begin{eqnarray}}
\def\en{\end{eqnarray}}
\def\non{\nonumber}
\def\la{\langle}
\def\ra{\rangle}
\def\pp{{\prime\prime}}

\def\vp{\varepsilon}

\def\ov{\overline}

\def\pr{{Phys. Rev.}~}
\def\prl{{ Phys. Rev. Lett.}~}
\def\pl{{ Phys. Lett.}~}
\def\np{{ Nucl. Phys.}~}

\def\lsim{ {\ \lower-1.2pt\vbox{\hbox{\rlap{$<$}\lower5pt\vbox{\hbox{$\sim$}
}}}\ } }
\def\gsim{ {\ \lower-1.2pt\vbox{\hbox{\rlap{$>$}\lower5pt\vbox{\hbox{$\sim$}
}}}\ } }

\begin{document}

\title{Covariant Light-Front Approach for $s$-wave
and $p$-wave Mesons}

\author{Chun-Khiang Chua}
%
\affiliation{Institute of Physics, Academia Sinica, Taipei, Taiwan
115, Republic of China }

\date{\today}

\begin{abstract}
We study the decay constants and form factors of the ground-state
$s$-wave and low-lying $p$-wave mesons within a covariant
light-front approach. Numerical results of the  $B\to D^{**}$
transition form factors, where $D^{**}$ denotes generically a
$p$-wave charmed meson, are compared with other model
calculations. Predictions on the decay rates for $\overline B\to
D^{**}\pi$, $D^{**}\rho$, $\overline D_s^{**} D^{(*)}$ by using
these decay constants and form factors are in agreement with the
experimental data. The universal Isgur-Wise functions,
$\xi,\,\tau_{1/2},\,\tau_{3/2}$ are obtained and are used to test
the Bjorken and Uraltsev sum rules.
\end{abstract}


\maketitle

\section{Introduction}

Mesonic weak transition form factors and decay constants are two
of the most important ingredients in the study of hadronic weak
decays of mesons. There exist many different model calculations.
The light-front quark model \cite{Ter,Chung} is the only
relativistic quark model in which a consistent and fully
relativistic treatment of quark spins and the center-of-mass
motion can be carried out. This model is very suitable to study
hadronic form factors. Especially, as the recoil momentum
increases (corresponding to a decreasing $q^2$), we have to start
considering relativistic effects seriously. In particular, at the
maximum recoil point $q^2=0$ where the final-state meson could be
highly relativistic, there is no reason to expect that the
non-relativistic quark model is still applicable.

The relativistic quark model in the light-front approach has been
employed to obtain decay constants and weak form
factors~\cite{Jaus90,Ji92,Jaus96,Cheng97}. There exist, however,
some ambiguities and even some inconsistencies in extracting the
physical quantities. Well known examples are the vector decay
constant $f_V$, the form factor $F_0(q^2)$ in the pseudoscalar to
pseudoscalar transition \cite{Jaus99} and the electromagnetic form
factor $F_2(q^2)$ of the vector meson (see e.g. \cite{Bakker02}).
A covariant model has been constructed in \cite{CCHZ} for heavy
mesons within the framework of heavy quark effective theory and
the results are free from above mentioned ambiguities.

\begin{figure}[t!]
\centerline{
            {\epsfxsize1.3 in \epsffile{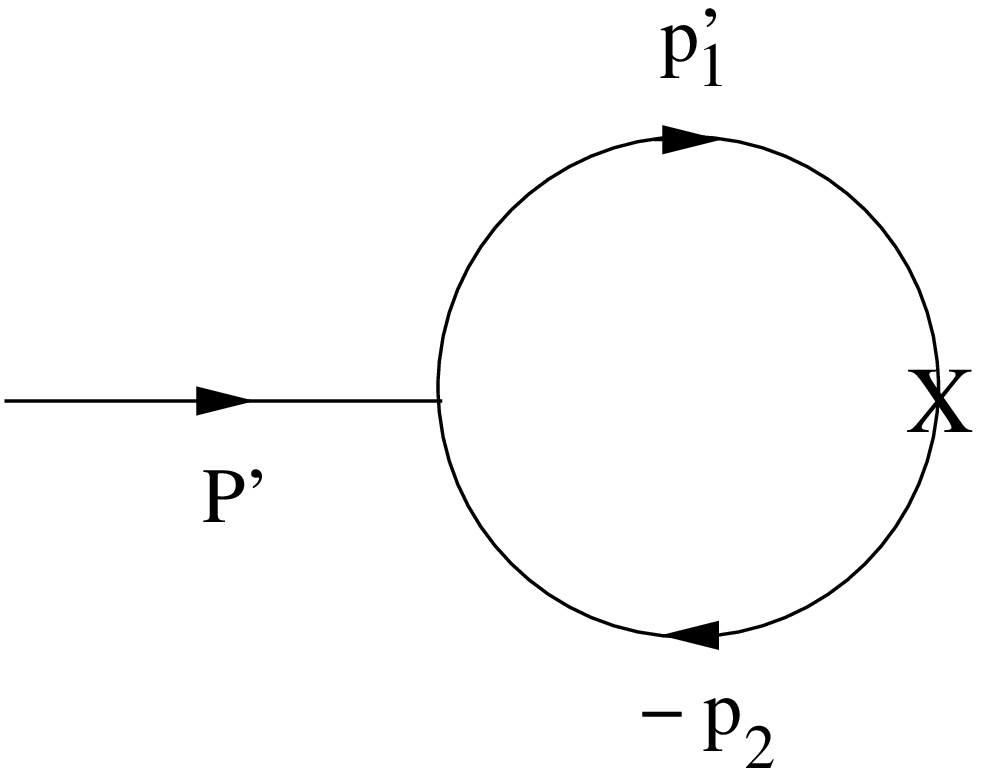}}
            \hspace{0.2cm}
            {\epsfxsize2.0 in \epsffile{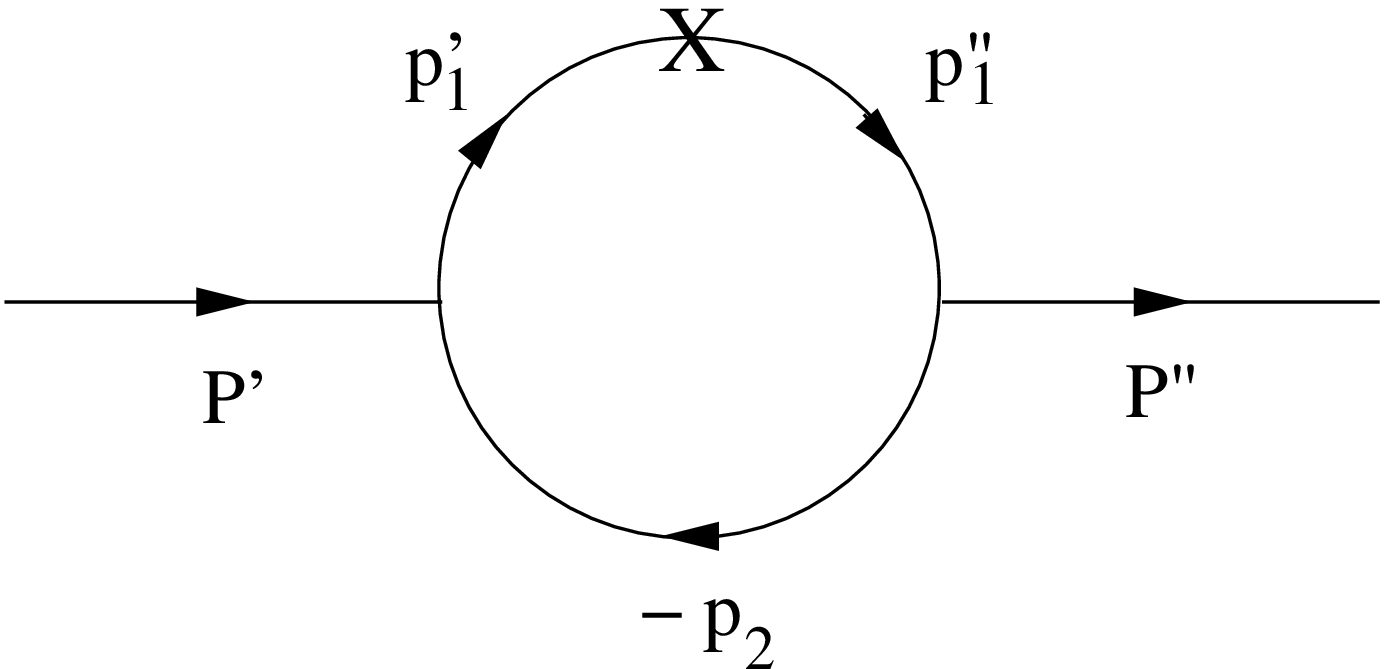}}}
\centerline{\,\,\,\,\,(a)\hspace{4.2cm}(b)} \vskip0.2cm
\caption{Feynman diagrams for (a) meson decay and (b) meson
transition amplitudes, where $P^{\prime(\pp)}$ is the incoming
(outgoing) meson momentum, $p^{\prime(\pp)}_1$ is the quark
momentum, $p_2$ is the anti-quark momentum and $X$ denotes the
corresponding $V-A$ current vertex.}\label{fig:feyn}
\end{figure}

Without appealing to the heavy quark limit, a covariant approach
of the light-front model for the usual pseudoscalar and vector
mesons has been put forward by Jaus \cite{Jaus99}. The procedure
of calculation can be separated into four steps: (a) The starting
point of the covariant approach is to consider the corresponding
covariant Feynman amplitudes in meson transitions as shown in
Fig.~\ref{fig:feyn}. (b) One can pass to the light-front approach
by using the light-front decomposition of the internal momentum in
covariant Feynman momentum loop integrals and integrating out the
$p^-=p^0-p^3$ component~\cite{CM69}. (c) At this stage one can
then apply some widely-used vertex functions in the conventional
light-front approach after $p^-$ integration. It is pointed out by
Jaus that in going from the manifestly covariant Feynman integral
to the light-front one, the latter is no longer covariant as it
receives additional spurious contributions proportional to the
lightlike vector $\tilde\omega^\mu=(1,0,0,-1)$. (d) This spurious
contribution is cancelled after correctly performing the
integration, namely, by the inclusion of the zero mode
contribution~\cite{zeromode}, so that the result is guaranteed to
be covariant. It should be noted that in \cite{Jaus99} a simple
covariant power-law-like vertex function is used in step~(a). Once
a covariant vertex function is used in step~(a), the above
procedure should give identical results to the direct integration
of Feynman integration via the usual technique. The power-law-like
vertex function does not lead to a satisfactory phenomenological
result when comparing to some widely used Gaussian form vertex
functions in step (c). Since the covariant counterpart of the
Gaussian like vertex function is not explicitly known, a use of it
in step (c) may lead to some residue spurious contributions. These
corrections to decay constants and form factors are worked out
in~\cite{Jaus03}. We check that these corrections are small in the
decay constant (within 10\%) and form factors (within 2\%).
In~\cite{CCH}, we have extended the covariant analysis of the
light-front model in~\cite{Jaus99} to even-parity, $p$-wave
mesons. Since the residue spurious corrections have not been
worked out in the $p$-wave meson case, we shall follow the
procedure of~\cite{Jaus99} for the extension.

Interest in even-parity charmed mesons has been revived by recent
discoveries of two narrow resonances: the $0^+$ state
$D_{s0}^*(2317)$~\cite{BaBar} and the $P^{1/2}_1$ state
$D_{s1}(2460)$~\cite{CLEO}, and two broad resonances,
$D_0^*(2308)$ and $D_1(2427)$~\cite{BelleD}. Furthermore, the
hadronic $B$ decays such as $B\to D^{**}\pi$ and $B\to D_s^{**}\ov
D$ have been recently observed, where $D^{**}$ denotes a $p$-wave
charmed meson. A theoretical study of them requires the
information of the $B\to D^{**}$ form factors and the decay
constants of $D^{**}$ and $D_s^{**}$. In the meantime, three body
decays of $B$ mesons have been recently studied at the $B$
factories: BaBar and Belle. The Dalitz plot analysis allows one to
see the structure of exclusive quasi-two-body intermediate states
in the three-body signals. The $p$-wave resonances observed in
three-body decays begin to emerge. Theoretically, the
Isgur-Scora-Grinstein-Wise (ISGW) quark model~\cite{ISGW} is so
far the only model in the literature that can provide a
systematical estimate of the transition of a ground-state $s$-wave
meson to a low-lying $p$-wave meson. However, this model and, in
fact, many other models in $P\to P,V$ ($P$: pseudoscalar meson,
$V$: vector meson) calculations, are based on the non-relativistic
constituent quark picture. As noted in passing, the final-state
meson at the maximum recoil point $q^2=0$ or in heavy-to-light
transitions could be highly relativistic. It is thus important to
consider a relativistic approach.

There are some theoretical constraints implied by heavy quark
symmetry (HQS) in the case of heavy-to-heavy transitions and
heavy-to-vacuum decays~\cite{IW89}. It is important to check if
the calculated form factors and decay constants do satisfy these
constraints. Furthermore, under HQS the number of the independent
form factors is reduced and they are related to some universal
Isgur-Wise (IW) functions. The relevant IW functions, namely,
$\xi$, $\tau_{1/2}$ and $\tau_{3/2}$ are obtained. One can then
study some properties of these IW functions, including the slopes
and sum rules~\cite{Uraltsev,Bjorken}.

\section{Decay constant}

The decay constants for $J=0,1$ mesons are defined by the matrix
elements
 \be \label{eq:AM}
  \la 0|A_\mu|P(P^\prime)\ra &=&i  f_P P^\prime_\mu ,\qquad
      \la 0|V_\mu|S(P^\prime)\ra= f_S P^\prime_\mu ,
  \non \\
  \la 0|V_\mu|V(P^\prime,\vp')\ra &=& M^\prime_V
  f_V\vp^\prime_\mu,
  \non \\
      \la 0|A_\mu|\,^{3(1)}\!A(P^\prime,\vp')\ra&=&
      M^\prime_{^3\!A(^1\!A)} f_{^3\!A(^1\!A)}\varepsilon^\prime_\mu, \non
 \en
where the $^{2S+1} L_J= {}^1S_0$, $^3P_0$, $^3S_1$, $^3P_1$,
$^1P_1$ and $^3P_2$ states of $q_1^\prime \bar q_2$ mesons are
denoted by $P$, $S$, $V$, $^3\!A$, $^1\!A$ and $T$, respectively.
Note that a $^3P_2$ state cannot be produced by a current. It is
useful to note that in the SU(N)-flavor limit ($m_1^\prime=m_2$)
we should have vanishing $f_S$ and $f_{^1\!A}$~\cite{Suzuki}.
These can be easily seen from charge conjugation. Under charge
conjugation $V_\mu\to -V_\mu$, while $A_\mu\to A_\mu$. For a
charge conjugated state, we have $C=(-)^{L+S}$. Thus, we must have
$f_S=f_{^1\!A}=0$ for these charge conjugated states. Through
SU(N) symmetry the above constraint should apply to all other
states in the same multiples.

Furthermore, in the heavy quark limit ($m_1^\prime\to\infty$), the
heavy quark spin $s_Q$ decouples from the other degrees of freedom
so that $s_Q$ and the total angular momentum of the light
antiquark $j$ are separately good quantum numbers. Hence, it is
more convenient to use the $L^j_J=P^{3/2}_2$, $P^{3/2}_1$,
$P^{1/2}_1$ and $P^{1/2}_0$ basis. It is obvious that the first
and the last of these states are $^3P_2$ and $^3P_0$,
respectively, while \cite{IW91}
 \be\label{eq:Phalf}
 \left|P^{3/2}_1\right\ra=\sqrt{\frac{2}{3}}\,\left|^1P_1\right\ra
 +{1\over \sqrt{3}}\,\left|^3P_1\right\ra,
 \non\\
 \left|P^{1/2}_1\right\ra={1\over \sqrt{3}}\,\left|^1P_1\right\ra
 -\sqrt{\frac{2}{3}}\,\left|^3P_1\right\ra. \en
Since, decay constants should be identical within each multiplet,
$(S^{1/2}_0,\,S^{1/2}_1),\,(P^{1/2}_0,\,P^{1/2}_1),\,(P^{3/2}_1,\,P^{3/2}_2)$,
heavy quark symmetry (HQS) requires~\cite{IW89,HQfrules}
\begin{equation} \label{eq:HQSf}
 f_V=f_P,\qquad
 f_{A^{1/2}}=f_S,\qquad
 f_{A^{3/2}}=0,
\end{equation}
where we have denoted the $P^{1/2}_1$ and $P^{3/2}_1$ states by
$A^{1/2}$ and $A^{3/2}$, respectively. It is important to check if
the calculated decay constants satisfy the non-trivial
SU(N)-flavor and HQS relations.

We now follow~\cite{Jaus99} to evaluate meson decay constants and
obtain
 \be
 f_P&=&\frac{N_c}{16\pi^3}\int dx_2 d^2p^\prime_\bot \frac{4h^\prime_P}{x_1
 x_2 (M^{\prime2}-M^{\prime2}_0)}(m_1^\prime x_2+m_2 x_1),
\non\\
 f_S&=&\frac{N_c}{16\pi^3}\int dx_2 d^2p^\prime_\bot \frac{4h^\prime_S}{x_1
 x_2 (M^{\prime2}-M^{\prime2}_0)}(m_1^\prime x_2-m_2 x_1),
 \non\\
 f_V&=&\frac{N_c}{4\pi^3}\int dx_2 d^2p^\prime_\bot \frac{h^\prime_V}{x_1
      x_2 (M^{\prime2}-M^{\prime2}_0)}
 \nonumber\\
      &&\times\left[x_1 M^{\prime2}_0-m_1^\prime (m_1^\prime-m_2)-p^{\prime2}_\bot
            +\frac{m_1^\prime+m_2}{w^\prime_V}\,p^{\prime2}_\bot \right],
 \non\\
 f_{^3\!A}&=&-\frac{N_c}{4\pi^3}\int dx_2 d^2p^\prime_\bot
            \frac{h^\prime_{^3\!A}}{x_1 x_2 (M^{\prime2}-M^{\prime2}_0)}
 \nonumber\\
      &&\times\left[x_1 M^{\prime2}_0-m_1^\prime (m_1^\prime+m_2)-p^{\prime2}_\bot
            -\frac{m_1^\prime-m_2}{w^\prime_{^3\!A}}\,p^{\prime2}_\bot \right],
 \nonumber\\
  f_{^1\!A}&=&\frac{N_c}{4\pi^3}\int dx_2 d^2p^\prime_\bot
            \frac{h^\prime_{^1\!A}}{x_1 x_2 (M^{\prime2}-M^{\prime2}_0)}
      \left(\frac{m_1^\prime-m_2}{w^\prime_{^1\!A}}\,p^{\prime2}_\bot \right),
 \non\\
 \label{eq:f}
 \en
where $M^\prime$ are meson masses, $m_1^\prime, m_2$ are quark
masses, $h^\prime$ are vertex functions, $M^\prime_0$ are kinetic
masses and $x_i$ are momentum fractions. Since $h^\prime$ and
$M^\prime_0$ are symmetric under the exchange of 1 and 2 in the
$m^\prime_1=m_2$ limit. It is clear that $f_S=f_{^1\!A}=0$ for
$m^\prime_1=m_2$. The SU(N)-flavor constraints on $f_S$ and
$f_{^1\!A}$ are thus satisfied.

\begin{table}[t!]
\caption{\label{tab:beta} The input parameter $\beta$ (in units of
GeV) in the Gaussian-type wave function.}
\begin{ruledtabular}
\begin{tabular}{|c|ccccc|}
$^{2S+1} L_J$
          & $\beta_{u\bar d}$
          & $\beta_{s\bar u}$
          & $\beta_{c\bar u}$
          & $\beta_{c\bar s}$
          & $\beta_{b\bar u}$
          \\
\hline $^1S_0$
          & $0.3102$
          & $0.3864$
          & $0.4496$
          & $0.4945$
          & $0.5329$
          \\
$^3P_0$
          & $\beta_{a_1}$
          & $\beta_{K(^3P_1)}$
          & $0.3305$
          & $0.3376$
          & $0.4253$
          \\
$^3S_1$
          & $0.2632$
          & $0.2727$
          & $0.3814$
          & $0.3932$
          & $0.4764$
          \\
$^3P_1$
          & $0.2983$
          & $0.303$
          & $0.3305$
          & $0.3376$
          & $0.4253$
          \\
$^1P_1$
          & $\beta_{a_1}$
          & $\beta_{K(^3P_1)}$
          & $0.3305$
          & $0.3376$
          & $0.4253$
          \\
\end{tabular}
\end{ruledtabular}
\end{table}

\begin{table}[t!]
\caption{\label{tab:f} Mesonic decay constants (in units of MeV)
obtained by using Eq.~(\ref{eq:f}). Those in parentheses are taken
as inputs to determine the corresponding $\beta$'s shown in Table
I.}
\begin{ruledtabular}
\begin{tabular}{|c|ccccc|}
$^{2S+1} L_J$
          & $f_{u\bar d}$
          & $f_{s\bar u}$
          & $f_{c\bar u}$
          & $f_{c\bar s}$
          & $f_{b\bar u}$
          \\
\hline $^1S_0$
          & $(131)$
          & $(160)$
          & $(200)$
          & $(230)$
          & $(180)$
          \\
$^3P_0$
          & $0$
          & $22$
          & $86$
          & $71$
          & $112$
          \\
$^3S_1$
          & $(216)$
          & $(210)$
          & $(220)$
          & $(230)$
          & $(180)$
          \\
$^3P_1$
          & $(-203)$
          & $-186$
          & $-127$
          & $-121$
          & $-123$
          \\
$^1P_1$
          & $0$
          & $11$
          & $45$
          & $38$
          & $68$
          \\
\hline $P^{1/2}_1$
          & --
          & --
          & $130$
          & $122$
          & $140$
          \\
$P^{3/2}_1$
          & --
          & --
          & $-36$
          & $-38$
          & $-15$
          \\
\end{tabular}
\end{ruledtabular}
\end{table}

In order to have numerical results for decay constants, we need to
specify the constituent quark masses and the parameter $\beta$
appearing in the Gaussian-type wave function. For constituent
quark masses (in units of GeV) we use
 $
m_{u,d}=0.26,\,\,m_s=0.37,\,\, m_c=1.40,\,\,m_b=4.64.
 $
Note that $m_s$ and $m_c$ are constrained from the measured
form-factor ratios in semileptonic $D\to K^*\ell\bar\nu$
decays~\cite{FOCUS}. Shown in Tables~\ref{tab:beta} and
\ref{tab:f} are the input parameter $\beta$ and decay constants,
respectively. In Table \ref{tab:f} the decay constants in
parentheses are used to determine $\beta$. For the purpose of an
estimation, for $p$-wave mesons in $D$, $D_s$ and $B$ systems we
shall use the $\beta$ parameters in the ISGW2 model~\cite{ISGW2}
up to some simple scaling. Two remarks are in order: (i) The
values of the parameter $\beta_V$ presented in Table
\ref{tab:beta} are slightly smaller than the ones obtained in the
earlier literature. It is interesting that $\beta_V$ in the ISGW2
model also have a similar reduction due to L-S interaction, which
is neglected in the original ISGW model in the mass spectrum
calculation. (ii) The $\beta$ parameters for $p$-wave states of
$D$, $D_s$ and $B$ systems are the smallest when compared to
$\beta_{P,V}$.

It is clear that the decay constant of light scalar resonances is
suppressed relative to that of the pseudoscalar mesons owing to
the small mass difference between the constituent quark masses.
However, as shown in Table~\ref{tab:f}, the suppression becomes
less effective for heavy scalar mesons because of heavy and light
quark mass imbalance.
The prediction of $f_S=21$ MeV for the scalar meson in the $s\bar
u$ content (see Table~\ref{tab:f}) is most likely designated for
the $K_0^*(1430)$ state. Notice that this prediction is slightly
smaller than the result of 42 MeV obtained in \cite{Maltman} based
on the finite-energy sum rules, and far less than the estimate of
$(70\pm10)$ MeV  in \cite{Chernyak}. It is worth remarking that
even if the light scalar mesons are made from 4 quarks, the decay
constants of the neutral scalars $\sigma(600)$, $f_0(980)$ and
$a_0^0(980)$ must vanish owing to charge conjugation invariance.

In principle, the decay constant of the scalar strange charmed
meson $D^*_{s0}$ can be determined from the hadronic decay $B\to
\ov DD_{s0}^*$ since it proceeds only via external $W$-emission.
Naively, it is expected that $D^*_{s0}$ has a slightly smaller
decay constant than that of $D^*_0$, in contrast to the
pseudoscalar case where $f_{D_s}>f_D$. However, a recent
measurement of the $D\bar D_{s0}^*$ production in $B$ decays by
Belle \cite{FPCP03} seems to indicate a $f_{D_{s0}^*}$ of order
$60\pm10$ MeV \cite{Cheng:2003id} which is close to the
expectation of 71 MeV (see Table \ref{tab:f}). The smallness of
this decay constant is due to the fact that comparing to the
$c\bar u$ system, the $c\bar s$ system is closer to the SU(N)
limit, where $f_S=0$. We will return to $B$ decays in the next
section.

Except for $a_1$ and $b_1$ mesons which cannot have mixing because
of the opposite $C$-parities, physical strange axial-vector mesons
are the mixture of $^3P_1$ and $^1P_1$ states, while the heavy
axial-vector resonances are the mixture of $P_1^{1/2}$ and
$P_1^{3/2}$. For example, $K_1(1270)$ and $K_1(1400)$ are the
mixture of $K_{^3P_1}$ and $K_{^1\!P_1}$ (denoted by $K_{1A}$ and
$K_{1B}$, respectively, by PDG \cite{PDG}) owing to the mass
difference of the strange and non-strange light quarks:
 \be \label{eq:K1mixing}
 K_1(1270)=K_{^3\!P_1} \sin\theta+K_{^1\!P_1}\cos\theta,
 \nonumber\\
 K_1(1400)=K_{^3\!P_1} \cos\theta-K_{^1\!P_1}\sin\theta,
 \en
with $\theta\approx -58^\circ$ as implied from the study of $D\to
K_1(1270)\pi,~K_1(1400)\pi$ decays \cite{Cheng:2003bn}. We use
$f_{K_1(1270)}=175$~MeV~\cite{Cheng:2003bn} to fix
$\beta_{K(^3P_1)}\simeq\beta_{K(^1P_1)}=0.303$~GeV and obtain
$f_{K_1(1400)}=-87$~MeV. Note that these $\beta_{K(^3P_1)}$,
$\beta_{K(^1P_1)}$ are close to $\beta_{K^*}$.

For $D$, $D_s$ and $B$ systems, it is clear from Table~\ref{tab:f}
that $|f_{A^{3/2}}|\ll f_S \lsim f_{A^{1/2}}$, in accordance with
the expectation from HQS [cf. Eq. (\ref{eq:HQSf})].
In fact, the HQS relations on decay constants Eq.~(\ref{eq:HQSf})
are verified in the HQ limit~\cite{CCH}.

\section{Form factors}

Form factors for $P\to P,V$ and $P$ to low-lying $p$-wave meson
transitions are defined by~\cite{BSW,ISGW}
 \begin{widetext}
  \begin{eqnarray} \label{eq:ffPPV}
 \la P(P^\pp)|V_\mu|P(P^\prime)\ra &=& \left(P_\mu-{M^{\prime 2}-M^{\pp 2}\over q^2}\,q_ \mu\right)
                                     F_1^{PP}(q^2)+{M^{\prime 2}-M^{\pp 2}\over q^2}\,q_\mu\,F_0^{PP}(q^2), \non \\
 \la V(P^\pp,\vp^\pp)|V_\mu|P(P^\prime)\ra &=& -{1\over
                                              M'+M^\pp}\,\epsilon_{\mu\nu\alpha \beta}\vp^{\pp*\nu}P^\alpha
                                              q^\beta  V^{PV}(q^2),
                                              \non\\
  \la V(P^\pp,\vp^\pp)|A_\mu|P(P^\prime)\ra &=& i\Big\{
                                               (M'+M^\pp)\vp^{\pp*}_\mu A_1^{PV}(q^2)-{\vp^{\pp*}\cdot P\over
                                               M'+M^\pp}\,P_\mu A_2^{PV}(q^2)
                                                 -2M^\pp\,{\vp^{\pp*}\cdot P\over
                                               q^2}\,q_\mu\big[A_3^{PV}(q^2)-A_0^{PV}(q^2)\big]\Big\},\non\\
 \la S(P^\pp)|A_\mu|P(P^\prime)\ra &=&
                                       -i\left[\left(P_\mu-{M^{\prime 2}-M^{\pp 2}\over q^2}\,q_
                                       \mu\right) F_1^{PS}(q^2)+{M^{\prime 2}-M^{\pp 2}\over q^2}
                                       \,q_\mu\,F_0^{PS}(q^2)\right], \non \\
 \la A(P^\pp,\vp^\pp)|V_\mu|P(P^\prime)\ra &=&
                                        -i\Bigg\{(m_P-m_A) \vp^*_\mu V_1^{PA}(q^2)  - {\vp^*\cdot
                                        P'\over m_P-m_A}P_\mu V_2^{PA}(q^2)
                                        - 2m_A {\vp^*\cdot P'\over
                                        q^2}q_\mu\left[V_3^{PA}(q^2)-V_0^{PA}(q^2)\right]\Bigg\},\non \\
 \la A(P^\pp,\vp^\pp)|A_\mu|P(P^\prime)\ra &=& -{1\over
                                        m_P-m_A}\,\epsilon_{\mu\nu\rho\sigma}\vp^{*\nu}P^\rho
                                        q^{\sigma}A^{PA}(q^2),\non\\
 \la T(P^\pp,\vp^\pp)|V_\mu|P(P^\prime)\ra
           &=& h(q^2)\epsilon_{\mu\nu\alpha\beta}\vp^{\pp*\nu\lambda}
                    P_\lambda P^\alpha q^\beta, \non \\
 \la T(P^\pp,\vp^\pp)|A_\mu|P(P^\prime)\ra
           &=& -i\Big\{k(q^2)\vp^{\pp*}_{\mu\nu}P^{\nu}+
                \vp^{\pp*}_{\alpha\beta}P^{\alpha} P^{\beta}
                [P_\mu b_+(q^2)+q_\mu b_-(q^2)]\Big\}.
 \end{eqnarray}
 \end{widetext}
with $P=P^\prime+p^{\pp}$, $q=p^\prime-p^\pp$,
$F_1^{PP}(0)=F_0^{PP}(0)$, $A_3^{PV}(0)=A_0^{PV}(0)$,
$V_3^{PA}(0)=V_0^{PA}(0)$, where
 \be
A_3^{PV}(q^2)=\,{M'+M^\pp\over
2M^\pp}\,A_1^{PV}(q^2)-{M'-M^\pp\over
2M^\pp}\,A_2^{PV}(q^2),\non\\
V_3^{PA}(q^2)=\,{m_P-m_A\over 2m_A}\,V_1^{PA}(q^2)-{m_P+m_A\over
2m_A}\,V_2^{PA}(q^2).
 \en
The definition here for dimensionless $P\to A$ transition form
factors differs than that in 
\cite{Cheng:2003id}
where the coefficients $(m_P\pm m_A)$ are replaced by $(m_P\mp
m_A)$.

We follow~\cite{Jaus99} to obtain $P\to P,V$ form factors and
extend the formalism to the $p$-wave meson case. The calculation
is done in the $q^+=0$ frame, where $q^2\leq 0$. We
follow~\cite{Jaus96} to analytically continue form factors to
timelike region.

To proceed we find that the momentum dependence of form factors in
the spacelike region can be well parameterized and reproduced in
the three-parameter form:
 \be \label{eq:FFpara}
 F(q^2)&=&\,{F(0)\over 1-a(q^2/m_{B}^2)+b(q^2/m_{B}^2)^2}\,\label{eq:FFpara}\\
 F(q^2)&=&\,{F(0)/(1-q^2/m_{B}^2)\over
 [1-a(q^2/m_{B}^2)+b(q^2/m_{B}^2)^2]},\label{eq:FFpara1}
 \en
for $B\to M$ transitions where the latter one is for the form
factor $V_2$ and the former one is for the rest. The parameters
$a$, $b$ and $F(0)$ are first determined in the spacelike region.
We then employ this parametrization to determine the physical form
factors at $q^2\geq 0$.

In Table~\ref{tab:LFBtoD} we show the form factors and their $q^2$
dependence for the $B$ to $D$, $D^*$, $D_0^*(2308)$, $D_1^{1/2}$,
$D_1^{3/2}$, $D_2^*(2460)$ transitions. Other results, including
$B(D)$ to $\pi$, $\rho$, $a_0(1450)$, $a_1(1260)$, $b_1(1235)$,
$a_2(1320)$, $K$, $K^*$, $K^*_0(1430)$, $K_{^1\!P_1}$,
$K_{^3\!P_1}$, $K_2^*(1430)$ transition form factors can be found
in \cite{CCH}.

In Table~IV, decay rates for $\overline B\to D^{**}\pi$,
$D^{**}\rho$, $\overline D_s^{**} D^{(*)}$ obtained in light-front
and ISGW2 models, respectively, are given. These are updated from
the analysis of~\cite{Cheng:2003id} by using the decay constants
and form factors shown in Tables~II and III,
respectively~\cite{update}. Since decay constants for $p$-wave
mesons are not provided in ISGW2 model, our decay constants and
their form factors are used for the ISGW2 results quoted.

\begin{table}[t]
\caption{Form factors for $B\to
D,\,D^*,D^*_0,D_1^{1/2},D_1^{3/2},D_2^*$ transitions are fitted to
the 3-parameter form Eq. (\ref{eq:FFpara}) except for the form
factor $V_2$ denoted by $^{*}$ for which the fit formula Eq.
(\ref{eq:FFpara1}) is used. For the purpose of comparison we quote
the result of ISGW2 in the lower half table.} \label{tab:LFBtoD}
\begin{ruledtabular}
\begin{tabular}{| c c c c || c c c c |}
~~~$F$~~~~~
    & $F(0)$~~~~~
    &$a$~~~~~
    & $b$~~~~~~
& ~~~ $F$~~~~~
    & $F(0)$~~~~~
    & $a$~~~~~
    & $b$~~~~~~
 \\
    \hline
$F^{BD}_1$
    & $0.67$
    & 1.25
    & 0.39
& $F^{BD}_0$
    & 0.67
    & 0.65
    & $0.00$ \\
$V^{BD^*}$
    & $0.75$
    & 1.29
    & 0.45
&$A^{BD^*}_0$
    & 0.64
    & 1.30
    & 0.31 \\
$A^{BD^*}_1$
    & 0.63
    & 0.65
    & 0.02
&$A^{BD^*}_2$
    & $0.61$
    & 1.14
    & 0.52
    \\
$F^{BD^*_0}_1$
    & $0.24$
    & 1.03
    & 0.27
& $F^{BD^*_0}_0$
    & 0.24
    & $-0.49$
    & 0.35 \\
$A^{BD^{1/2}_1}$
    & $-0.12$
    & 0.71
    & 0.18
&$V^{BD^{1/2}_1}_0$
    & 0.08
    & 1.28
    & $-0.29$
    \\
$V^{BD^{1/2}_1}_1$
    & $-0.19$
    & $-1.25$
    & 0.97
& $V^{BD^{1/2}_1}_2$
    & $-0.12$
    & 0.67
    & 0.20
   \\
$A^{BD^{3/2}_1}$
    & $0.23$
    & 1.17
    & 0.39
&$V^{BD^{3/2}_1}_0$
    & $0.47$
    &  1.17
    &  0.03
    \\
$V^{BD^{3/2}_1}_1$
    & $0.55$
    & $-0.19$
    & 0.27
&$V^{BD^{3/2}_1}_2$
    & $-0.09^*$
    & $2.14^*$
    & $4.21^*$
    \\
$h$
    & 0.015
    & 1.67
    & 1.20
& $k$
    & 0.79
    & 1.29
    & 0.93
    \\
$b_+$
    & $-0.013$
    & 1.68
    & 0.98
& $b_-$
    & 0.011
    & 1.50
    & 0.91 \\
    \hline
$F_1^{BD_0^*}$
    & 0.18
    & 0.28
    & 0.25
& $F_0^{BD_0^*}$
    & 0.18
    & --
    & -- \\
 $A^{BD_1^{1/2}}$
    & $-0.16$
    & 0.87
    & 0.24
 &$V_0^{BD_1^{1/2}}$
    & $0.18$
    & 0.89
    & 0.25 \\
 $V_1^{BD_1^{1/2}}$
    & $-0.19$
    & --
    & --
 & $V_2^{BD_1^{1/2}}$ \
    & $-0.18$
    & 0.87
    & 0.24\\
 $A^{BD_1^{3/2}}$
    & $0.16$
    & 0.46
    & 0.065
 &$V_0^{BD_1^{3/2}}$
    & $0.43$
    & 0.54
    & 0.074 \\
 $V_1^{BD_1^{3/2}}$
    & $0.40$
    & $-0.60$
    & 1.15
 & $V_2^{BD_1^{3/2}}$
    & $-0.12$
    & 1.45
    & 0.83 \\
$h$
    & 0.011
    & 0.86
    & 0.23
& $k$
    & 0.60
    & 0.40
    & 0.68 \\
 $b_+$
    & $-0.010$
    & 0.86
    & 0.23
 & $b_-$
    & 0.010
    & 0.86
    & 0.23 \\
 \end{tabular}
\end{ruledtabular}
\end{table}

\begin{table}[t]
\caption{Updated decay rates for $\overline B\to D^{**}\pi$,
$D^{**}\rho$, $\overline D_s^{**} D^{(*)}$ obtained in light-front
and ISGW2 models respectively~\cite{Cheng:2003id}. Since decay
constants for $p$-wave mesons are not provided in ISGW2 model, we
use our decay constants and their form factors for the ISGW2
results quoted below.}
\begin{ruledtabular}
\begin{tabular}{|l c c l l| }
${\mathcal B}(10^{-3})$
         & LF
         & ISGW2
         & Expt
         & Ref
         \\
\hline
$B^-\to D^*_0(2308)^0\pi^-$
         & $0.83$
         & $0.52$
         & $0.92\pm0.29$
         & Belle
         \\
 $B^-\to D_1(2427)^0\pi^-$
         & $ 0.52$
         & $1.1$
         & $0.75\pm0.17$
         & Belle
         \\
 $B^-\to D'_1(2420)^0\pi^-$
         & $1.3$
         & $1.0$
         & $1.0\pm0.2$
         & Belle,BaBar
         \\
         &
         &
         & $1.5\pm0.6$
         & PDG
         \\
 $B^-\to D^*_2(2460)^0\pi^-$
         & $1.2$
         & $0.66$
         & $0.78\pm0.14$
         & Belle,BaBar
         \\
 $B^-\to D^*_0(2308)^0\rho^-$
         & $1.7$
         & $1.0$
         &
         &
         \\
 $B^-\to D_1(2427)^0\rho^-$
         & $1.1$
         & $2.1$
         &
         &
         \\
 $B^-\to D'_1(2420)^0\rho^-$
         & $3.7$
         & $2.6$
         & $<1.4$
         & PDG
         \\
 $B^-\to D^*_2(2460)^0\rho^-$
         & $3.4$
         & $1.8$
         & $<4.7$
         & PDG
         \\
 \hline
 $B^-\to \ov D^*_{s0}(2317)^-D^0$
         & $1.3$
         & --
         & $0.85\pm0.33$
         & Belle
         \\
 $B^-\to \ov D_{s1}(2460)^-D^0$
         & $1.9$
         &  --
         & $1.5\sim 4.4$
         & Belle
         \\
 $B^-\to \ov D'_{s1}(2536)^-D^0$
         & $0.55$
         &--
         &
         &
         \\
 $B^-\to \ov D^*_{s0}(2317)^-D^{*0}$
         & $0.70$
         &--
         &
         &
         \\
 $B^-\to \ov D_{s1}(2460)^-D^{*0}$
         & $7.1$
         &--
         &
         &
         \\
 $B^-\to \ov D'_{s1}(2536)^-D^{*0}$
         & $1.4$
         & --
         &
         &
         \\
\end{tabular}
\end{ruledtabular}
\end{table}

Several remarks are in order:
 \begin{enumerate}
 \item
For heavy-to-heavy transitions such as $B\to D$, $D^*$, $D^{**}$,
the sign of various form factors can be checked by heavy quark
symmetry. Our results are indeed in accordance with HQS.

 \item
It is pointed out in \cite{Cheng97} that for $B\to D,D^*$
transitions, the form factors $F_1,A_0,A_2,V$ exhibit a dipole
behavior, while $F_0$ and $A_1$ show a monopole dependence.
An inspection of Table~\ref{tab:LFBtoD} indicates that form
factors $F^{BD}_0$ and $A_1^{BD^*}$ have a monopole behavior,
while $F_1^{BD}$, $V^{BD^*}$ and $A^{BD_1^{3/2}}$ have a dipole
dependence.

\item Our numerical result for $k$ is too sensitive to the $\beta$
parameter. The $k$ shown in Table~III are determined from $h$ and
$b_+-b_-$ through HQ relations instead.

 \item
We see from the comparison of LF and ISGW2 results in
Table~\ref{tab:LFBtoD} that the form factors at small $q^2$
obtained in the covariant light-front and ISGW2 models differ not
more than 40$\%$. Relativistic effects are mild in $B\to D$
transition, but they could be more prominent in heavy to light
transitions, especially at maximum recoil ($q^2=0$). For example,
we obtain $V_0^{Ba_1}(0)=0.13$~\cite{CCH}, while ISGW2 gives 1.01.
If $a_1(1260)$ behaves as the scalar partner of the $\rho$ meson,
it is expected that $V_0^{Ba_1}\sim A_0^{B\rho}\sim O(0.1)$.

 \item To determine the physical form factors for $B\to
D_1(2427),D_1(2420),D_{s1}(2460),D_{s1}(2536)$ transitions, one
needs to know the mixing angles of $D_1^{1/2}-D^{3/2}_1$. A mixing
angle $\theta_{D_1}=(5.7\pm2.4)^\circ$ is obtained by Belle
through a detailed $B\to D^*\pi\pi$ analysis \cite{BelleD}, while
$\theta_{D_{s1}}\approx 7^\circ$ is determined from the quark
potential model \cite{Cheng:2003id} as the present upper limits on
the widths of $D_{s1}(2460)$ and $D'_{s1}(2536)$ do not provide
any constraints on the $D_{s1}^{1/2}\!-\!D_{s1}^{3/2}$ mixing
angle. We use a theoretical predicted $\theta_{D_1}=12^\circ$ in
Table~IV~\cite{Cheng:2003id}.

\item The decay rates for $\overline B\to D^{**}\pi$,
$D^{**}\rho$, $\overline D_s^{**} D^{(*)}$ obtained in light-front
and ISGW2 models shown in Table~IV agree with experimental results
in most cases. (i)~Note that our decay constants for $p$-wave
mesons are used in both LF and ISGW2 cases. (ii)~Usually we expect
a factor two to three enhancement in $D^{**}\rho$ rates from
$D^{**}\pi$ rates. The old upper limit in $D^\prime_1(2420)\rho$
needs further check. (iii)~The color-allowed Cabbibo favored
$\overline B\to \overline D_s^{**} D^{(*)}$ amplitudes involve
$\overline B\to D^{(*)}$ form factors and $D_s^{**}$ decay
constants. Since the form factors are standard, this type of
decays provides valuable information on $D_s^{**}$ decay
constants. The agreement between theory and experiment supports
our predictions on $D_s^{**}$ decay constants.

\end{enumerate}

\section{Isgur-Wise functions}

In \cite{CCH} Isgur-Wise functions are obtained through either top
down~\cite{CCHZ} or bottom up (take $m_Q\to \infty$) approaches.
The IW functions can be fitted nicely to the form
 \be
 \xi(\omega) &=& 1-1.22(\omega-1)+0.85(\omega-1)^2, \non \\
 \tau_{1/2}(\omega) &=& 0.31\left(1-1.18(\omega-1)+0.87(\omega-1)^2\right), \non \\
 \tau_{3/2}(\omega) &=&
 0.61\left(1-1.73(\omega-1)+1.46(\omega-1)^2\right),
 \en
where we have used the same $\beta_\infty$ parameter for both
initial and final wave functions.
Our results are similar to that obtained in the ISGW model
\cite{ISGW} (numerical results for the latter being quoted from
\cite{Morenas}). Our result $\rho^2=1.22$ for the slope parameter
is consistent with the current world average of $1.44\pm0.14$
extracted from exclusive semileptoic $B$ decays \cite{HFAG}.

%
%

It is interesting to notice that there are Uraltsev and Bjorken
sum rules~\cite{Uraltsev,Bjorken}
 \be
 \sum_n|\tau_{3/2}^{(n)}(1)|^2-\sum_n|\tau_{1/2}^{(n)}(1)|^2={1\over 4}\,,
 \non\\
\rho^2={1\over
 4}+\sum_n|\tau_{1/2}^{(n)}(1)|^2+2\sum_n|\tau_{3/2}^{(n)}(1)|^2\,,
 \en
respectively, where $n$ stands for radial excitations and $\rho^2$
is the slope of the IW function $\xi(\omega)$. The Bjorken and
Uraltsev sum rules for the Isgur-Wise functions are fairly
satisfied.

\section{Conclusions}

We have studied the decay constants and form factors of the
ground-state $s$-wave and low-lying $p$-wave mesons within a
covariant light-front approach. Our main results are as follows:
(i) The SU(N) and HQ relations on decay constants are satisfied.
(ii) The decay constant of scalar mesons is suppressed relative to
that of the pseudoscalar mesons. The smallness of the decay
constant of the newly observed $D_{s0}^*(2317)$ implied by a
recent Belle measurement on $B\to \ov DD_{s0}^*$ could be
accommodated. (ii) Numerical results of the form factors for $B\to
D,~D^*,~D^{**}$ transitions are presented in detail. At $q^2=0$
our results are close to ISGW2 results within 40~\%. (iii) The
prediced decay rates for $\overline B\to D^{**}\pi$, $D^{**}\rho$,
$\overline D_s^{**} D^{(*)}$ obtained in light-front and ISGW2
models shown in Table~IV agree with experimental results. (iv) The
universal Isgur-Wise functions $\xi(\omega),\tau_{1/2}(\omega)$
and $\tau_{3/2}(\omega)$ are obtained. The Bjorken and Uraltsev
sum rules for the Isgur-Wise functions are fairly satisfied.

\acknowledgments I am grateful to my collaborators Hai-Yang Cheng
and Chien-Wen Hwang. This research was supported 
by the National Science Council of R.O.C. under Grant Nos.
NSC92-2811-M-001-054.

\end{document}